\documentclass[fleqn,10pt]{wlscirep}
\usepackage{geometry}
\usepackage[utf8]{inputenc}
\usepackage[numbers]{natbib}
\usepackage{amssymb}
\usepackage{amsmath} 
\usepackage{siunitx}
\usepackage{todonotes}
\usepackage{filecontents}

\newcommand{\figref}[1]{Fig. \ref{#1}}
\newcommand{\f}[1]{\mathcal{F}\left[#1\right]}
\newcommand{\finv}[1]{\mathcal{F}^{-1} \left[#1\right]}
\newcommand{\norm}[2]{\left\lVert#1\right\rVert_#2}

\title{Low-dose cryo electron ptychography via non-convex Bayesian optimization}

\author[1,2,3,*]{Philipp Michael Pelz}
\author[4]{Wen Xuan Qiu}
\author[1,2]{Robert Buecker}
\author[1,2]{Guenther Kassier}
\author[1,2,3,4]{R.J. Dwayne Miller}

\affil[1]{Max Planck Institute for the Structure and Dynamics of Matter, 22761 Hamburg, Germany}
\affil[2]{Center for Free Electron Laser Science, Luruper Chaussee 149, 22761 Hamburg, Germany}
\affil[3]{Department of Physics, University of Hamburg, Hamburg 22761, Germany}
\affil[4]{Departments of Chemistry and Physics, University of Toronto, 80 St. George Street, Toronto M5S 1H6, Canada}
\affil[*]{philipp.pelz@mpsd.mpg.de}

\begin{abstract}
	Electron ptychography has seen a recent surge of interest for phase sensitive imaging at atomic or near-atomic resolution. However, applications are so far mainly limited to radiation-hard samples because the required doses are too high for imaging biological samples at high resolution.
	We propose the use of non-convex, Bayesian optimization to overcome this problem and reduce the dose required for successful reconstruction by two orders of magnitude compared to previous experiments. We suggest to use this method for imaging single biological macromolecules at cryogenic temperatures and demonstrate 2D single-particle reconstructions from simulated data with a resolution of \SI{7.9}{\angstrom} at a dose of \SI{20}{\elementarycharge^-\per\angstrom^2}. When averaging over only 15 low-dose datasets, a resolution of \SI{4}{\angstrom} is possible for large macromolecular complexes. With its independence from microscope transfer function, direct recovery of phase contrast and better scaling of signal-to-noise ratio, cryo-electron ptychography may become a promising alternative to Zernike phase-contrast microscopy.
\end{abstract}

\begin{document}
	
	\flushbottom
	\maketitle
	
	\section*{Introduction}
\begin{figure}[h!]
		\includegraphics[width=1\textwidth]{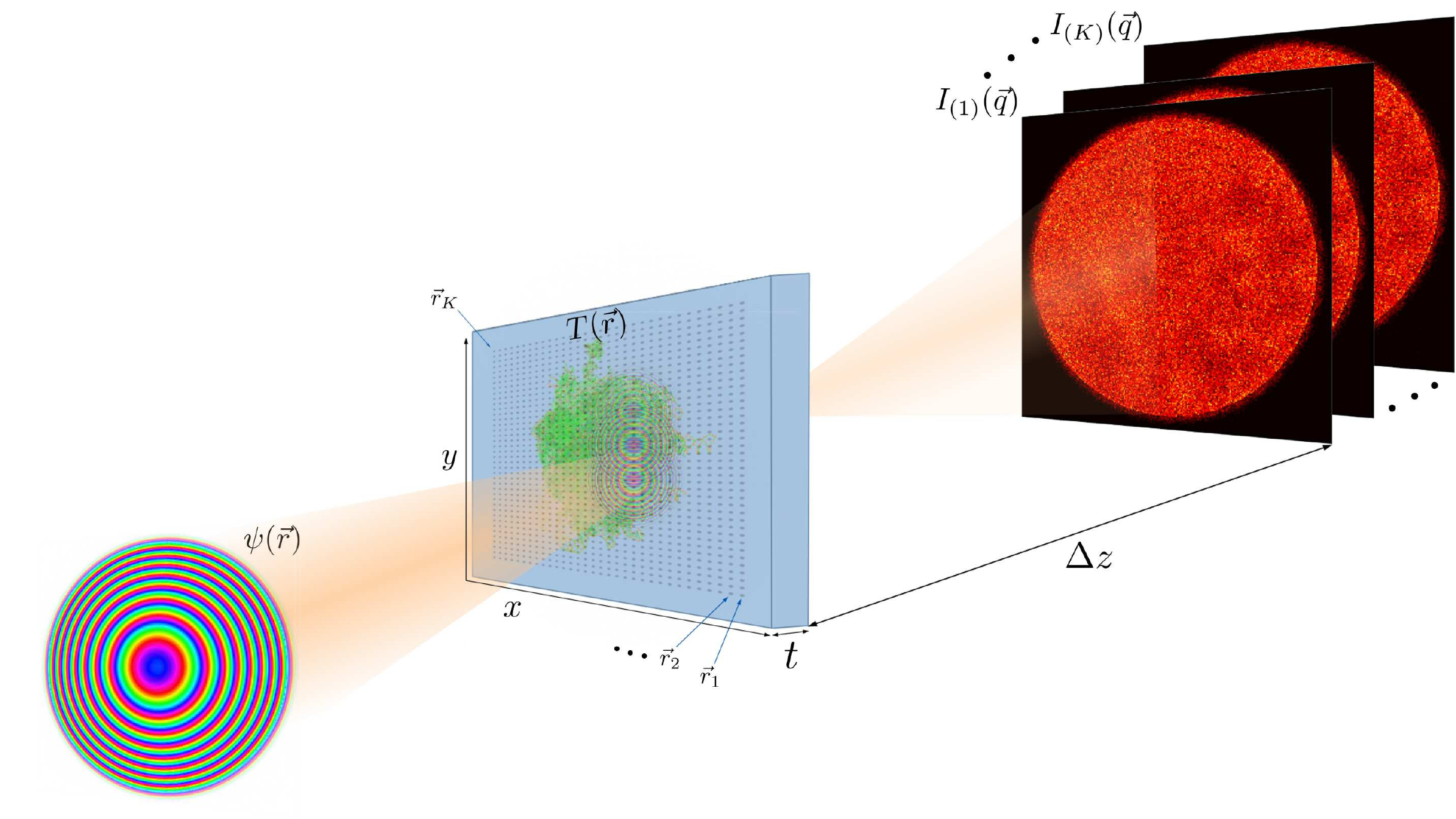}
		\caption{Experimental geometry in ptychography. The coherent electron wave function $\psi(\vec{r})$ illuminates several regions of the sample, characterized by the transmission function $T(\vec{r})$. For every position, a 2D diffraction pattern is recorded in the far field. The sample thickness $t$ can be neglected for biological macromolecules in the reconstruction at the resolutions presented in this paper.}
		\label{fig:experiment}
	\end{figure}

	The advent of direct electron detectors has led to a resolution revolution in the field of cryo-electron microscopy in the last few years, producing three-dimensional atomic potential maps of biological macromolecules of a few \SI{100}{\kilo\dalton} with a resolution better than \SI{3.5}{\angstrom} \cite{bartesaghi_structure_2014,groll_structure_1997}, so that individual amino acid side-chains can be resolved. An important part in this revolution is new image processing algorithms based on a Bayesian approach, which infer important parameters without user intervention \cite{scheres_relion:_2012} and correct beam induced motion \cite{bai_ribosome_2013}. However, several challenges remain to be overcome to routinely reach \SI{3}{\angstrom} resolution also for small complexes \cite{subramaniam_cryoem_2016,bai_how_2015}: Firstly, beam-induced specimen charging and subsequent motion currently still render the high resolution information of the first few frames of a high repetition rate movie recorded with a direct electron detector unusable \cite{scheres_beam-induced_2014}, because the motion is too fast to efficiently correct for it. Secondly, the Detective Quantum Efficiency (DQE) of detectors can still be improved at high spatial frequencies \cite{glaeser_how_2016,bai_how_2015}. Thirdly, the contrast of single images can still be improved to enable reconstructions with fewer particles and increase the throughput \cite{glaeser_how_2016}.\\
	The last of these challenges has recently been addressed with a new phase plate model \cite{danev_cryo-em_2016,danev_volta_2014}, which is comparatively simple to use and provides excellent contrast at low spatial frequencies. In addition to this hardware-based approach to achieve linear phase contrast in the measured amplitudes, discovered by Zernike in the 1930s \cite{zernike_how_1955}, it is also possible to algorithmically retrieve the phase information from a set of coherent diffraction measurements. One such technique, commonly known as \textit{ptychography} or \textit{scanning coherent diffractive microscopy} \cite{Thibault2008}, is becoming increasingly popular in the field of materials science due to experimental robustness and the possibility to obtain quantitative phase contrast with an essentially unlimited field of view \cite{maiden_quantitative_2015,Diaz2012}. The use of ptychography for imaging radiation sensitive samples with electrons at high resolution is however precluded so far by its high dose requirements.\\
	We show here how the use of non-convex Bayesian optimization to solve the ptychographic phase retrieval problem fulfills the dose requirements for imaging biological macromolecules and makes it possible to obtain 2D images from single particles with sub-nanometer resolution. After a short introduction into the technique, we will also mention how ptychography offers improvements for the other two challenges discussed above.\\
	Despite being first proposed as a solution to the phase problem for electrons \cite{hoppe_trace_1982,rodenburg_phase_1989}, ptychography has seen its biggest success in X-ray imaging, due to the less stringent sample requirements and the experimental need for lensless imaging techniques. Recent developments include the introduction of iterative algorithms to enable the reconstruction of datasets collected with an out-of focus probe \cite{Thibault2009,Maiden2009}, which decreases the memory requirements of the method dramatically; the algorithmic corrections of experimental difficulties such as unknown scan positions \cite{guizar-sicairos_phase_2008,zhang_translation_2013,maiden_annealing_2012}, partial coherence \cite{Thibault2012}, probe movement during exposure \cite{pelz_--fly_2014,Clark2014}, intensity fluctuations during the scan \cite{Thibault2012,marchesini_augmented_2013} and reconstruction of background noise \cite{marchesini_augmented_2013,maiden_quantitative_2015}.\\
	In recent years, some of these advances have been applied in the context of electron microscopy and yielded atomic resolution reconstructions of low-atomic number materials \cite{Putkunz2012,yang_simultaneous_2016,dalfonso_dose-dependent_2016} and quantitative phase information \cite{maiden_quantitative_2015}.\\
	\figref{fig:experiment} shows the experimental set-up for an out-of focus ptychography experiment. A ptychographic dataset is collected by scanning a spatially confined, coherent beam, subsequently called 'probe', over the specimen and recording far-field diffraction patterns at a series of positions such that the illuminated regions of neighboring positions overlap. The diffraction limited resolution of the final image is given by the maximum angle subtended by the detector $\mathrm{r_d = \frac{\lambda * \Delta z}{N_{pix}/2*d_{pix}}}$, where $N_{pix}$ is the number of detector pixels, $d_{pix}$ is the detector pixel size, $\lambda$ is the wavelength. Given the set of positions and a realistic forward model which describes the formation of the diffraction pattern, the complex-valued transmission function which describes the atomic properties of the specimen \cite{lubk_phase-space_2015}, can be retrieved by solving a non-convex inverse problem. 
	The electron dose used for successful reconstruction has been above \SI{1e3}{\elementarycharge^-\per\angstrom^2} so far, limiting its usability to radiation-hard specimens. Table \ref{tab:title} lists recently published electron ptychography experiments and the used doses.
	The lowest dose was reported in \cite{Humphry2012}, which used an estimated \SI{3.33e3}{\elementarycharge^-\per\angstrom^2} at a resolution of \SI{58.4}{\pico\metre}, resulting in a dose of \SI{1.1e3}{\elementarycharge^-\per pixel} and achieving a line resolution of \SI{2.3}{\angstrom}, demonstrated by resolving the lattice spacing of gold nanoparticles.
	We demonstrate here via simulations that it is possible to reduce the dose by almost a factor of \num{100} to reach the dose range required for imaging biological macromolecules.\\
	The problem of beam-induced sample movement has already been addressed before the development of fast direct detectors. Scanning with small spots of several \SI{10}{\nano\meter} in size over a vitrified sample has shown to reduce beam induced specimen movement \cite{bullough_use_1987,brink_computer-controlled_1992,downing_spot-scan_1991} in real-space imaging and it has been noted that the remaining movement may be due to radiation damage, not sample charging \cite{bullough_use_1987}. Ptychography naturally operates with a confined beam, thus minimizing the area where charge can build up, such that the movement should be reduced compared to the illumination of large areas in cryo-EM. Additionally we note that due to ptychography being a scanning technique, fast acquisition and rastering is instrumental to achieve high throughput. This and the sampling requirements given by the experimental setup allow to operate the detector with very large effective pixel sizes, such that the DQE and MTF are at near-perfect values.
	\section*{Results}
	\subsection*{Image formation of cryo-TEM and cryo-ptychography}
	We perform multislice simulations of three different biological macromolecules with molecular weights ranging from \SI{64}{\kilo\dalton} to \SI{4}{\mega\dalton}. We choose the \SI{64}{\kilo\dalton} hemoglobin \cite{fermi_crystal_1984}, the \SI{706}{\kilo\dalton} 20S proteasome from yeast \cite{groll_structure_1997}, and the \SI{4}{\mega\dalton} human ribosome \cite{anger_structures_2013}.We create atomic potential maps using the Matlab code InSilicoTem \cite{vulovic_image_2013} with a thickness of \SI{70}{\nano\meter} at an electron energy of \SI{300}{\kilo\electronvolt}. We use the isolated atom superposition approximation, without solving the Poisson-Boltzmann equations for the interaction between the molecule and the ions. We also do not model the amorphousness of the solvent, which was performed in \cite{vulovic_image_2013} using molecular dynamics simulations, but was seen to have a negligible effect at very low doses. As described in \cite{vulovic_image_2013}, we model the imaginary part of the potential via the inelastic mean free path, creating a minimal transmission contrast between the vitreous ice and the protein. Using these potential maps, we simulate a ptychography experiment by cropping three-dimensional slices from this potential at several positions and propagate a coherent incoming wave through the potential slices using the methods described in \cite{Kirkland2010} in custom code. We model the detector properties as in \cite{vulovic_image_2013}, by multiplying the Fourier transform of the exit-wave intensity with $\sqrt{DQE(\vec{q})}$ before applying shot noise. Subsequently we convolve with the noise transfer function of the detector to yield the final intensity.\\
	A notable difference in the simulations and in practice is the fact that for cryo-EM usually no binning is applied to maximize the imaged area and increase throughput. Therefore, the also high spatial-frequency regions with low values of the DQE and NTF are used for image formation. For ptychography on the other hand the detector can be heavily binned until the scanning beam still fits into the real-space patch given by $\mathrm{r_d \cdot N_{pix}}$. We therefore apply a 14x binning to the diffraction patterns and crop to a size of \num{256x256}, i.e. we only use the detector up to \SI{6.7}{\percent} Nyquist frequency. The real-space patch covered by the detector is then \SI{51}{\nano\meter} at for a $\mathrm{r_d}=$\SI{2}{\angstrom}, more than necessary to fit a \SI{10}{\nano\meter} size beam. This leads to a near-constant DQE and a near unity NTF, so that there is no necessity to include them in the ptychography reconstructions, while we still include them in the simulation of the diffraction data. We note, however that a convolution with a detector transfer function can be modeled with a partially coherent beam if necessary, as demonstrated in \cite{Thibault2013b,Enders2014}.
	We choose the Gatan K2 Summit as the detector for our simulations because it has the highest published DQE and MTF values at low spatial frequencies at \SI{300}{\kilo\electronvolt} \cite{mcmullan_comparison_2014}. We note that other direct detection cameras with faster readout may be more suitable for ptychographic scanning experiments \cite{ryll_pnccd-based_2016,mcmullan_detective_2009}, but characteristics for these cameras at \SI{300}{\kilo\electronvolt} are either not published or worse than the K2 Summit. Additionally, the use of the K2 Summit for both ptychography and phase-contrast TEM simulations simplifies a direct comparison between the two methods.\\
	The final model for the formation of the intensity on the detector is then 
	\begin{equation}
	I_0(\vec{q}) = |\f{\psi_{exit}(\vec{r})}|^2 
	\end{equation}
	for the diffraction pattern and 
	\begin{equation}
	I_0(\vec{q}) = |\finv{\f{\psi_{exit}(\vec{r})}\cdot\mathrm{CTF}(\vec{q})}|^2 \\
	\end{equation}
	for the cryo-EM image. The intensity after detection is modeled as 
	\begin{equation}
	I(\vec{q}) =\finv{\f{\mathrm{Poisson}\left(\finv{\f{I_0(\vec{q})}\cdot\sqrt{\mathrm{DQE}(\vec{q})}}\right)}\cdot\mathrm{NTF}(\vec{q})}\, , 
	\end{equation}
	where NTF is the noise transfer function of the detector \cite{meyer_characterisation_2000}.
	\subsection*{Single-particle reconstructions}
\begin{figure}[h!]
		\includegraphics[width=1\textwidth]{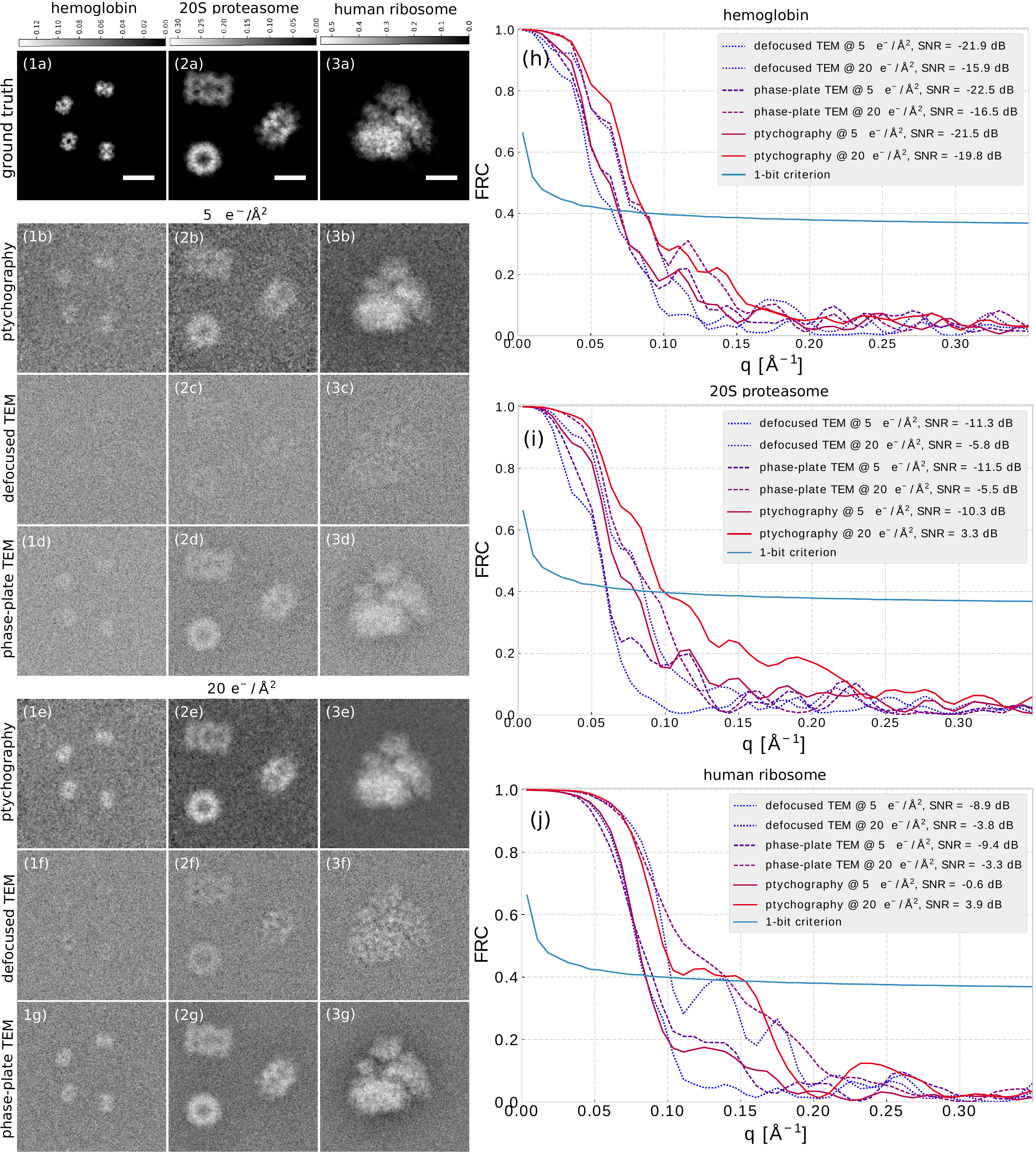}
		\caption{Cryo-electron ptychography reconstructions from simulated data and simulated cryo-EM images for different doses and 3 macromolecules with growing molecular weights in columns 1-3. Row a): Phase of the transmission function, the ground truth for the ptychography reconstructions. The scale bar above the figures is in rad. Rows b) and e): ptychography reconstruction at doses of \SI{5}{\elementarycharge^-\per\angstrom^2} and \SI{20}{\elementarycharge^-\per\angstrom^2}. Rows c) and f): Simulated cryo-EM image with a defocus of \SI{1.6}{\micro\meter} at a dose of \SI{5}{\elementarycharge^-\per\angstrom^2} and \SI{20}{\elementarycharge^-\per\angstrom^2}. Rows d) and g): Simulated cryo-EM image with a Zernike phase plate and a defocus of \SI{50}{\nano\meter} at doses of \SI{5}{\elementarycharge^-\per\angstrom^2} and \SI{20}{\elementarycharge^-\per\angstrom^2}. Column $\mathrm{(1)}$ hemoglobin, column $\mathrm{(2)}$ 20S proteasome, column $\mathrm{(3)}$ human ribosome. The scale bar is \SI{10}{\nano\meter}. Signal-to-noise ratio over spatial frequency for h) hemoglobin, i) 20S proteasome, j) human ribosome.}
		\label{fig:reconstructions}
	\end{figure}

	Fig. \ref{fig:reconstructions} shows a comparison of low-dose ptychography reconstructions with currently used methods for single-particle imaging with electrons: defocus-based cryo-EM and Zernike phase contrast cryo-EM with a Volta phase-plate. We choose exemplary doses of \SI{5}{\elementarycharge^-\per\angstrom^2} as the typical threshold where the highest resolution details are destroyed \cite{stark_electron_1996} and \SI{20}{\elementarycharge^-\per\angstrom^2} as a typical dose at which experiments are performed. We have reversed the contrast in the cryo-EM images to simplify the visual comparison with the ptychography reconstructions. It can be seen that ptychography clearly produces the best images for the larger particles at both doses, while it gives roughly the same results as Zernike phase-contrast for very small particles like hemoglobin. To quantitatively assess the image quality, we have computed both the Fourier Shell Correlation (FRC) \cite{Heel2005} and the average signal-to-noise ratio (SNR) in Fig. \ref{fig:reconstructions} h)-i). As ground truth for the cryo-EM images we use the electron counts in a noiseless, aberration-free image. We choose here the 1-bit criterion as a resolution threshold, as the point where the SNR drops to 1 \cite{Heel2005}. With this criterion we achieve resolutions of \SI{14.6}{\angstrom} and \SI{11.5}{\angstrom} for hemoglobin at \SI{5}{\elementarycharge^-\per\angstrom^2} and \SI{20}{\elementarycharge^-\per\angstrom^2}. For 20S proteasome \SI{12.7}{\angstrom} and \SI{10.1}{\angstrom}; and for human ribosome \SI{11.5}{\angstrom} and \SI{7.9}{\angstrom} respectively at doses of \SI{5}{\elementarycharge^-\per\angstrom^2} and \SI{20}{\elementarycharge^-\per\angstrom^2}. Looking at the average SNR values, our ptychography approach seems to have the biggest advantages for particles with molecular weight of several \SI{100}{\kilo\dalton} and larger. Here, the improvements over conventional methods are \SI{8.8}{\decibel} for 20S proteasome at \SI{20}{\elementarycharge^-\per\angstrom^2} and \SI{7.2}{\decibel} for human ribosome at \SI{20}{\elementarycharge^-\per\angstrom^2}, almost an order of magnitude improvement. For the ribosome, the signal is strong enough that the oscillating contrast transfer of the defocus-based method is nicely visible in the FRC.
	\subsection*{Effect of averaging}
\begin{figure}[h!]
		\includegraphics[width=1\textwidth]{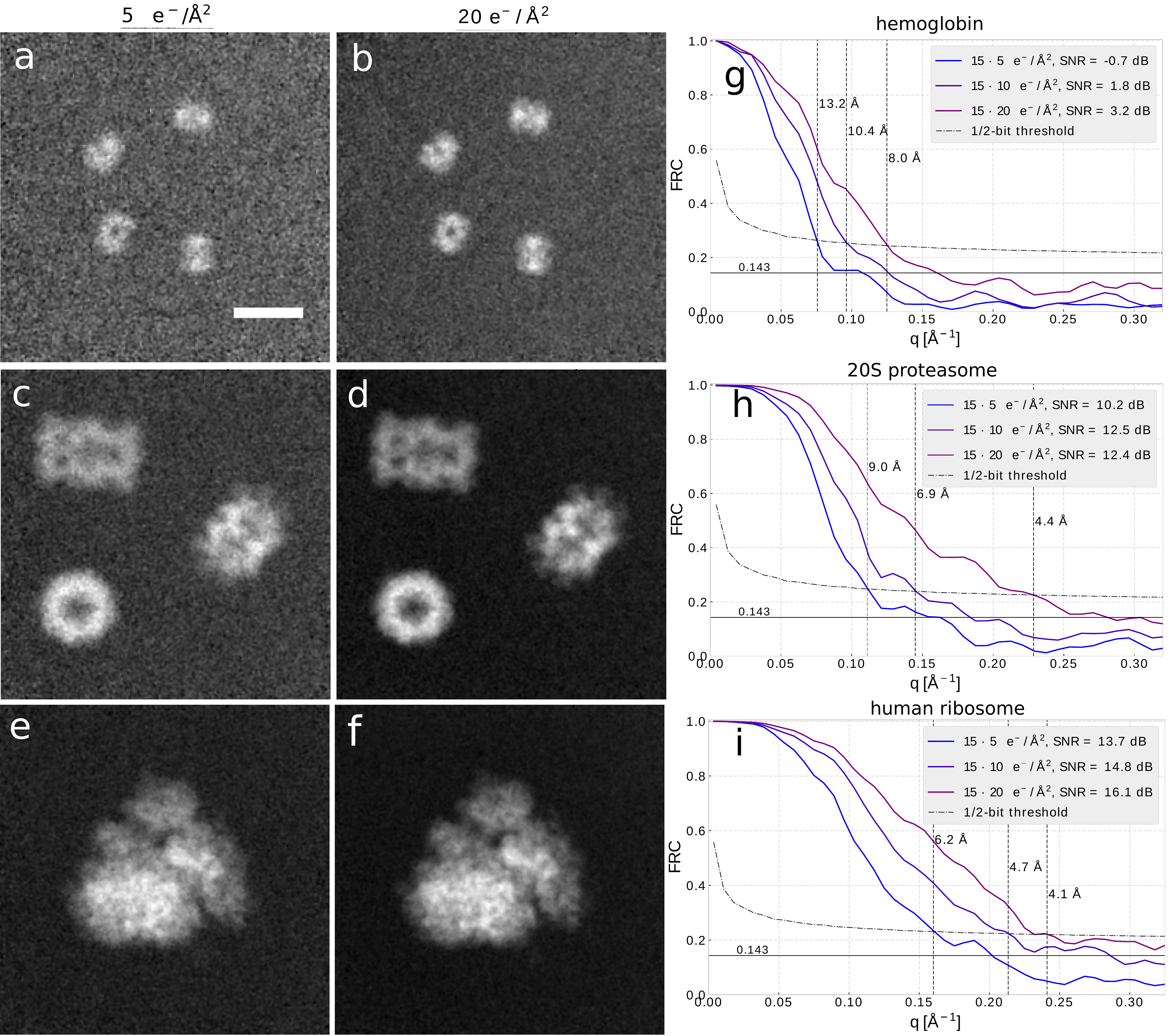}
		\caption{Average over 15 ptychographic reconstructions from independent data sets for a) hemoglobin with \SI{5}{\elementarycharge^-\per\angstrom^2}, b) hemoglobin with \SI{20}{\elementarycharge^-\per\angstrom^2}, c) proteasome 20S with \SI{5}{\elementarycharge^-\per\angstrom^2}, d) proteasome 20S with \SI{20}{\elementarycharge^-\per\angstrom^2}, e) human ribosome with \SI{5}{\elementarycharge^-\per\angstrom^2}, f) human ribosome with \SI{20}{\elementarycharge^-\per\angstrom^2}. FRC of averaged reconstructions from independent data sets for g) hemoglobin, h) proteasome 20S i) human ribosome.}
		\label{fig:averaging}
	\end{figure}

	In single-particle 3D cryo-EM, a large ensemble of 2D images is collected and then oriented and averaged in three dimensions to improve the signal-to-noise ratio. A 3D reconstruction from ptychographic data is out of the scope of this paper, because dedicated algorithms need to be developed to achieve optimal results. The reconstructed 2D images do not obey the same noise statistics as cryo-EM micrographs and therefore the use of standard software would lead to suboptimal results. In the best case, a 3D model can be reconstructed directly from the raw diffraction data, while coarse orientation alignment could be done in real space from 2D reconstructions shown here.\\
	To give a rough estimate how the resolution and SNR of our algorithm scales with averaging multiple datasets, we perform here an computer experiment and average over reconstructions of 15 datasets where the particles are in identical orientation. We choose an average of 15 because there is no significant improvement when averaging more datasets. Fig. \ref{fig:averaging} a)-f) shows images of the averaged reconstructions of our three samples, at \SI{5}{\elementarycharge^-\per\angstrom^2} and \SI{20}{\elementarycharge^-\per\angstrom^2} respectively. Comparing the SNR values with the single-particle reconstructions, we see for almost all doses and samples more than an order of magnitude increase in SNR. For hemoglobin, the average SNR increases from \SI{-16.8}{\decibel} to \SI{-0.7}{\decibel} at \SI{5}{\elementarycharge^-\per\angstrom^2} and from \SI{-14}{\decibel} to \SI{3.2}{\decibel} at \SI{20}{\elementarycharge^-\per\angstrom^2}. For the 20S proteasome the SNR increases from \SI{-5.5}{\decibel} to \SI{10}{\decibel} at \SI{5}{\elementarycharge^-\per\angstrom^2} and from \SI{6.2}{\decibel} to \SI{12}{\decibel} at \SI{20}{\elementarycharge^-\per\angstrom^2}. For the human ribosome the SNR increases from \SI{-0.6}{\decibel} to \SI{13.7}{\decibel} at \SI{5}{\elementarycharge^-\per\angstrom^2} and from \SI{3.9}{\decibel} to \SI{16.1}{\decibel} at \SI{20}{\elementarycharge^-\per\angstrom^2}. All in all averaging of reconstructions gives larger improvements for very low doses.\\
	We also FRC curves for the averaged reconstructions of two independently created data sets to give a resolution estimate. We use here the 1/2-bit resolution threshold discussed in \cite{Heel2005}, which gives a slightly more conservative estimate than the 0.143-criterion commonly used in averaged reconstructions for cryo-EM. With averaging, a resolution of \SI{8}{\angstrom} is achieved for hemoglobin, \SI{4.4}{\angstrom} is achieved for 20S proteasome and \SI{3.9}{\angstrom} for human ribosome at a dose of \SI{20}{\elementarycharge^-\per\angstrom^2}.
	\subsection*{Probe and dose dependence}
\begin{figure}[h!]
		\includegraphics[width=1\textwidth]{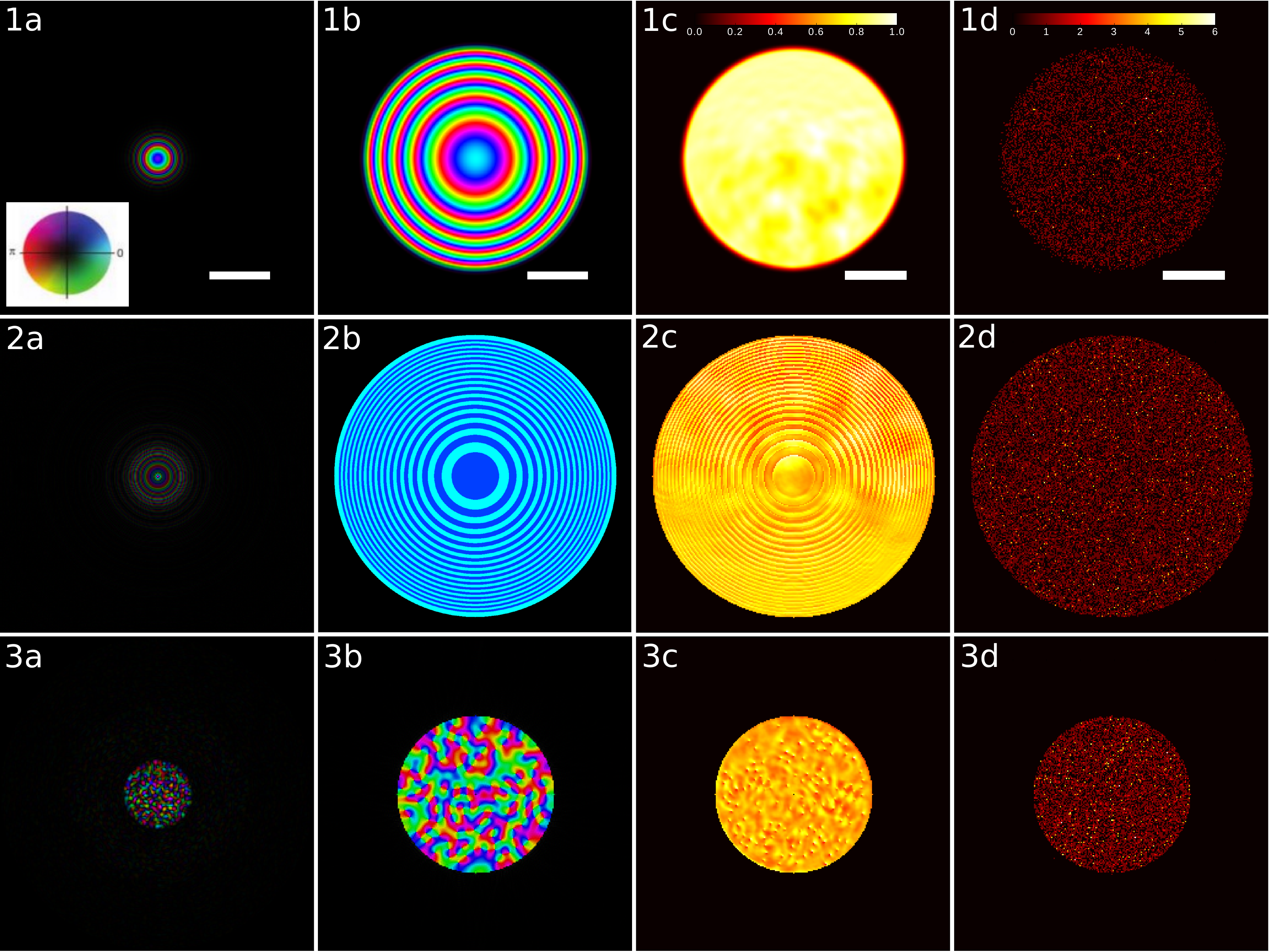}
		\caption{Different probes evaluated in this paper and corresponding diffraction patterns. Row 1: defocused beam with defocus aberration of 400nm, convergence half-angle \SI{7.85}{\milli\radian}; row 2: defocused beam created by a Fresnel zone plate, \SI{600}{\nano\meter} from focus; row 3: randomized beam, generated by a holographic phase plate and focused by a conventional lens. Column a) beam in real space, at the sample position, scale bar is \SI{10}{\nano\meter}; column b) beam at the probe forming aperture, scale bar is \SI{3.85}{\milli\radian}; column c) diffraction pattern of human ribosome at unlimited dose, normalized to the maximum intensity; column d) diffraction pattern for a scan with 589 acquisitions, at an electron dose of \SI{20}{\elementarycharge^-\per\angstrom^2}. The inset in 1a shows the color wheel that is used to represent amplitude and phase in columns a) and b).}
		\label{fig:probes}
	\end{figure}

\begin{figure}[h!]
		\includegraphics[width=1\textwidth]{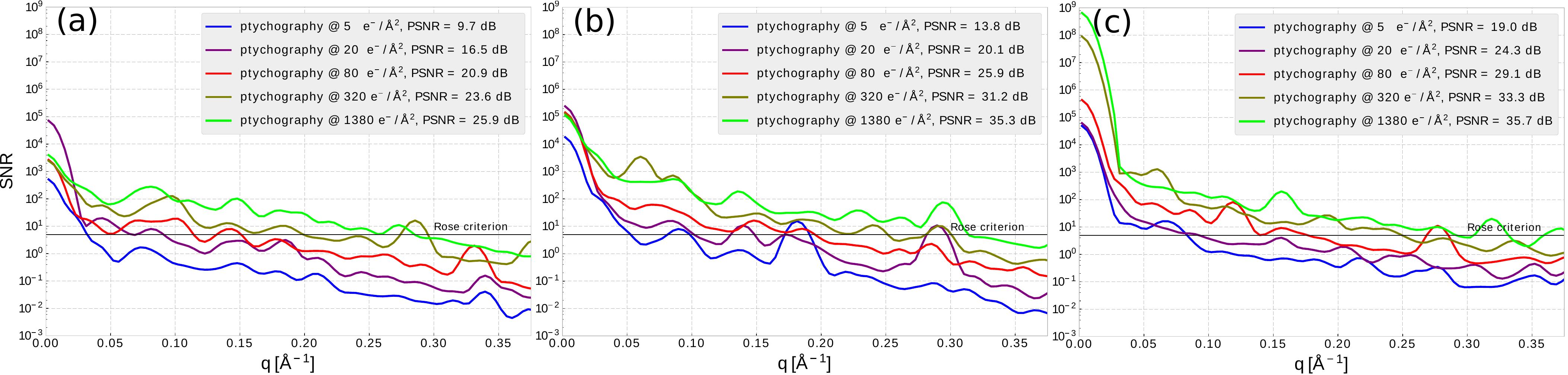}
		\caption{Signal to noise ratios of reconstructions of the human ribosome at different radiation doses using a) the defocused probe, b) the Fresnel zone plate, c) the randomized probe}
		\label{fig:probe_snr}
	\end{figure}

	It is well known that the phase profile of the ptychographic probe can heavily influence the reconstruction quality \cite{Marchesin2014,li_multiple_2016,maiden_soft_2013,guizar-sicairos_role_2012,li_ptychographic_2014}. Here we numerically test three different probes, depicted in Fig. \ref{fig:probes}, and their influence on the reconstruction SNR at low and high doses: 1) a standard defocused probe with defocus aberration of 400 nm, 2) a defocused Fresnel Zone Plate (FZP), and 3) a randomized probe generated by a holographic phase plate and a focusing lens. Fig. \ref{fig:probes} depicts these probe in real and Fourier space and typical diffraction patterns at infinite dose and low dose. The FZP was recently suggested as a phase modulator for bright-field STEM \cite{ophus_efficient_2016}, because its simple phase modulation allows analytical retrieval of linear phase contrast. However, diffractive optics typically have imperfections due to the manufacturing process which introduce errors and dose inefficiency if the phase modulation is obtained by a simple fitting procedure. Iterative ptychography algorithms allow for the simultaneous retrieval of the probe wave function \cite{Maiden2009, Thibault2009} and therefore offers full flexibility in the design of the phase profile. Empirically, probes with a diffuse phase profile result in better reconstructions, therefore we test as a third probe a random illumination generated by a holographic phase plate and a focusing lens.\\
	Fig. \ref{fig:probe_snr} shows the SNR of the three proposed probes as a function of spatial frequency. It can be seen that at very low doses of \SI{5}{\elementarycharge^-\per\angstrom^2}, the randomized probe achieves the best SNR. At high doses of \SI{1380}{\elementarycharge^-\per\angstrom^2}, the SNR of the randomized probe at very low frequencies is higher up to a factor of \num{1e4} than the defocused probe and the FZP, whereas the performance at high resolution is only slightly better. We give two qualitative explanations of this fact, but emphasize that a theory for optimal measurement design in ptychography and a practically feasible implementation of it is still outstanding and may drastically improve upon the results presented here.
	\newpage
	\section*{Methods}
	\subsection*{Mathematical framework of Ptychography}
	The two-dimensional complex transmission function of the object is discretized as a $\mathrm{n_1 \times n_2}$ matrix and denoted as $T : D^{n_1 \times n_2}_{r_d} \rightarrow \mathbb{C}$, where $r_\mathrm{d} > 0$ is the diffraction limited length scale. The object is illuminated by a small beam with known distribution, and discretized as a $\mathrm{m_1 \times m_2}$ matrix denoted as $\psi : D^{M}_{r_d} \rightarrow \mathbb{C}$. For simplicity, in this paper we only consider the case  $\mathrm{n_1 = n_2}$ and $\mathrm{m_1 = m_2}$, i.e. a uniform discretization in both axes.
	In the experiment, the beam is moved over the sample to positions $\vec{r}_i$ and illuminates $\mathrm{K > 1}$ subregions to obtain K diffraction images.
	\begin{equation}
	\mathbf{I}_i(\vec{q}) = |\mathcal{F}\left[\psi(\vec{r}+\vec{r}_{i}) \cdot T(\vec{r})\right]|^2, i \in \{0,...,K\}\\,
	\end{equation}
	where the discretized real space coordinates are discretized in steps of $r_\mathrm{d}$  and reciprocal space coordinates are discretized in steps of $\frac{1}{m_i\cdot r_d}$.
	Mathematically, ptychography can be understood as a special case of the \textit{generalized phase retrieval problem}: given a phase-less vector of measurements $\mathbf{y} \in \mathbb{R}^m_+$
	find a complex vector $\mathbf{z} \in \mathbb{C}^n$ such that 
	\begin{equation}
	\mathbf{y} = |\mathcal{A}\mathbf{z}|^2,
	\end{equation}
	where $\mathcal{A} : \mathbb{C}^n \rightarrow \mathbb{C}^m$ is an arbitrary linear operator. We follow here the notations in \cite{Marchesin2014} to formulate ptychography as a generalized phase retrieval problem. First, we vectorize the transmission function as $\mathbf{T}^V \in \mathbb{C}^N$ with $\mathrm{N = n_1 \cdot n_2 \in \mathbb{N}}$. We introduce the matrix $R_{(i)} \in \mathbb{R}^{M\times N}$, which extracts an $m_1 \times m_2$ sized area centered at position $\vec{r}_i$ from T. With these notations in place, the relationship between the noise-free diffraction measurements collected in a ptychography experiment and $\mathbf{T}^V$ can be represented compactly as
	\begin{eqnarray}
	\mathbf{I} &=& |\mathbf{F}\mathbf{Q}\mathbf{T}^V|^2 = |\mathbf{P}\mathbf{T}^V|^2, \\
	\end{eqnarray}
	where $\mathbf{P}$ is constructed by cropping K regions from T, multiplying by the incoming beam, and a 2D discrete Fourier transform $\mathbf{F}$, i.e. $\mathbf{P}=\mathbf{F}\mathbf{Q}$.
	\begin{eqnarray}
	\overbrace{
		\begin{bmatrix}
		I_{1} \\
		\vdots \\
		I_{KM}
		\end{bmatrix}}^{\mathbf{I}\in\mathbb{R}^{KM}}&=&\left|\overbrace{\begin{bmatrix}
		F & \cdots & 0 \\
		\vdots & \ddots & \vdots \\
		0 & \cdots & F
		\end{bmatrix}}^{\mathbf{F}\in\mathbb{C}^{KM \times KM}}\overbrace{\begin{bmatrix}
		\text{diag}(\psi) R_{(1)} \\
		\vdots \\
		\text{diag}(\psi) R_{(K)}
		\end{bmatrix}}^{\mathbf{Q}\in\mathbb{C}^{KM \times N}}\mathbf{T}^V\right|^2\\.
	\end{eqnarray}
	The matrix $\mathbf{P} \in \mathbb{C}^{KM \times N}$ is sometimes called design matrix because its entries determine the measurement outcome and reflect the experimental design. In the last decades many algorithms to solve this problem have been devised, of which we only review a small fraction  with regards to low-dose reconstruction in the following section. For the subsequent analysis, we denote the KM row vectors of \textbf{P} as $\mathbf{p}_i$.
	\subsection*{Bayesian optimization with truncated gradients}\label{sec:algos}
	The most prominent iterative algorithms to solve the ptychographic phase retrieval problems are the difference map (DM) \cite{Thibault2008} algorithm and the extended ptychographic iterative engine (ePIE) \cite{Maiden2009}. The difference map belongs to the family of algorithms which use projections onto non-convex sets to reach a fix-point, the solution lying at the intersection of the two sets. It can be shown that the standard algorithm of alternating projections is equivalent to steepest descent optimization with a Gaussian likelihood and is not suited for low-dose reconstructions \cite{Marchesin2014} because the Poisson distribution arising from discretized count events differs too strongly from a Gaussian at very low electron counts. While this equality does not hold for the more elaborate projection algorithms like DM and relaxed averaged alternating reflections (RAAR) \cite{luke_relaxed_2005}, they also fail in practice at low doses \cite{Godard2012,pelz_--fly_2014}, and statistical reconstruction methods have to be used. Thibault and Guizar-Sicairos \cite{Thibault2012} have analyzed maximum likelihood methods in conjunction with a conjugate gradient update rule as a refinement step, after the DM algorithm has converged. They demonstrate improved SNR compared to the DM algorithm alone. They note, however, that starting directly with maximum likelihood optimization often poses convergence problems.\\
	The PIE algorithm can be formulated as maximum-a-posteriori (MAP) optimization with a Gaussian likelihood function and an independently weighted Gaussian prior of the object change \cite{Godard2012,Thibault2013b}, combined with a stochastic gradient-like update rule. The noise performance of the PIE algorithm has been investigated in \cite{Godard2012} and found to be worse than the Poissonian likelihood model at low counting statistics.
	While practically very robust, both algorithms can get stuck in local minima and until recently \cite{Marchesin2014}, no proof of convergence to a global minimum existed. \cite{Godard2012} suggest to use a global gradient update at the start to avoid stagnation, and \cite{maiden_quantitative_2015} use a restarted version with the stochastic gradient update rule, after removal of phase vortex artifacts.\\
	Due to the lack of algorithms with provable converge guarantees, the mathematical community has recently picked up the problem and a host of new algorithms with provable convergence has been developed, which we do not elaborate on here but point the interested reader to the summary articles \cite{jaganathan_phase_2015,shechtman_phase_2015} and the article \cite{sun_geometric_2016-1}, which refers to the most recent developments.\\
	Here we focus on developments which specifically target low-dose applications. Notable in this area is the work by Katkovnik et al. \cite{Katkovnik2013a}, which in addition to the maximum likelihood estimate introduces a transform-domain sparsity constraint on the object and optimizes two objective functions in an alternating fashion: one for the maximizing the likelihood and one for obtaining a sparse representation of the transmission function. However, instead of including the Poissonian likelihood directly, an observation filtering step is performed with a Gaussian likelihood. To obtain a sparse representation of the object, the popular BM3D denoising filter is used. Another recent, similar approach uses dictionary learning to obtain a sparse representation of the transmission function \cite{tillmann_dolphin_2016-1}, however only real-valued signals are treated.
	During the writing of this paper, Yang et al. suggested using the one-step inversion technique for low-dose ptychography \cite{yang_enhanced_????}, however no statistical treatment of the measurement process is included so far, which could be a promising avenue for future research.
	We formulate ptychographic phase retrieval as a Bayesian inference problem by introducing the probability of the transmission function $\mathbf{T}^V$, given a set of measurements $\mathbf{y} = \left(y_1, \, y_2, \, ... \, ,y_{KM}\right)^T \in \mathbb{R}_{+}^{KM}$ with the Bayes' rule:
	\begin{equation}
	P(\mathbf{T}^V|\mathbf{y}) = \frac{P(\mathbf{y}|\mathbf{T}^V)\,P(\mathbf{T}^V)}{P(\mathbf{y})}
	\end{equation}
	Since the measurements $y_i$ follow the Poisson distribution
	\begin{equation}
	y_i \sim \mathrm{Poisson}(I_i(\mathbf{T}^V)),
	\end{equation}
	the Likelihood function is given by 
	\begin{equation}
	P(y_i|\mathbf{T}^V) = \frac{I_i(\mathbf{T}^V)^{y_i}}{y_i!}e^{-I_i(\mathbf{T}^V)}.
	\end{equation}
	The prior distribution is usually chosen such that it favors realistic solutions, so that noise is suppressed in the reconstructed image. Here we evaluate two different models. A simple prior, suggested in \cite{Thibault2013b}, penalizes large gradients in the image with a Gaussian distribution on the gradient of the transmission function, which is also known as Tikhonov regularization.
	\begin{equation}
	P_{Tikhonov}(T) = \exp{(-\frac{\mu_0}{\kappa}||\nabla T(\vec{r})||^2)} = \exp{(-\frac{\mu_0}{\kappa}\sum_{i=1}^N (\mathrm{D}_x\mathbf{T}^V)_i^2 + (\mathrm{D}_y\mathbf{T}^V)_i^2)}
	\end{equation}
	with $\kappa = 8\frac{N^2}{N_m ||I||_1}$ chosen as in \cite{Thibault2013b} to scale the numerical value of the prior to be close to the likelihood. $\mathrm{D}_x$ and $\mathrm{D}_y$ are the discrete forward difference operators. The second prior we evaluate is based on the work by Katkovnik et al. \cite{Katkovnik2013a} and uses sparse modeling to denoise the transmission function:
	\begin{equation}
	P_{sparse}(\mathbf{T}^V) = \exp{(- \mu || \mathbf{T}^V - \mathbf{T}^V_{sparse} ||^2)}
	\end{equation}
	Here,  $\mathbf{T}^V_{sparse}$ is built up by applying the BM3D collaborative filtering algorithm \cite{danielyan_bm3d_2012-1}, which we describe here shortly. The BM3D algorithm computes the transform-domain sparse representation in 4 steps: 1$)$ itfinds the image patches similar to a given image patch and groups them in a 3D block 2$)$ 3D linear transform of the 3D block; 3$)$ shrinking of the transform spectrum coefficients; 4$)$ inverse 3D transformation. As input for the BM3D algorithm we transform $\mathbf{T}^V$ into hue-saturation-value format, the phase representing hue and the amplitude representing value, with full saturation. The prior $P_{sparse}(T)$ reduces the difference between the denoised version of the current transmission function and the transmission function itself. We note that during the writing of this manuscript an extensive evaluation of denoising phase retrieval algorithms was published \cite{chang_general_2016}, which also evaluates BM3D denoising using an algorithm similar to the one presented here and finds superior performance compared to other denoising strategies such as total variation or nonlocal means.
	We do not take into account the marginal likelihood $P(\mathbf{y})$ due to the high dimensionality of the problem. 
	Given the likelihood function and the prior distribution, one can write down the objective function for the MAP estimate:
	\begin{equation}
	\mathbf{T}^V_{MAP} := \underset{\mathbf{T}^V}{\mathrm{arg min}}\,\,\mathcal{L}_{MAP}(\mathbf{T}^V)
	\end{equation}
	The gradient of the likelihood is given as 
	\begin{equation}
	\mathcal{L}(\mathbf{T}^V) = \sum_{i=1}^{KM}\left[|\mathbf{p}_i\, \mathbf{T}^V|^2- y_i\, \mathrm{log}(|\mathbf{p}_i\, \mathbf{T}^V|^2)\right].
	\label{equ:LMAP1}
	\end{equation}
	and the MAP objective functions are
	\begin{equation}
	\mathcal{L}_{MAP}(\mathbf{T}^V) = -\log\left(\frac{P(\mathbf{y}|\mathbf{T}^V)P(\mathbf{T}^V)}{P(\mathbf{y})}\right) =  \mathcal{L}(\mathbf{T}^V) + \frac{\mu_0}{\kappa}||\nabla T(\vec{r})||^2
	\label{equ:LMAP2}
	\end{equation}
	and 
	\begin{equation}
	\mathcal{L}_{BM3D-MAP}(\mathbf{T}^V) = \mathcal{L}(\mathbf{T}^V) + \mu_1 || \mathbf{T}^V - \mathbf{T}^V_{sparse} ||^2
	\label{equ:LMAP3}
	\end{equation}
	We calculate the gradient of $\mathcal{L}_{MAP}(T)$
	\begin{equation}
	\nabla\mathcal{L}_{MAP}(T) = \sum^{KM}_{i=1} 2 \, \mathbf{p}_i\, \mathbf{T}^V \left(1-\frac{y_i }{|\mathbf{p}_i\, \mathbf{T}^V|^2}\right)\mathbf{p}_i^{\dagger} +  2 \frac{\mu_0}{\kappa}\sum_{i=1}^N (\mathrm{D}_x\mathbf{T}^V)_i + (\mathrm{D}_y\mathbf{T}^V)_i
	\end{equation}
	\begin{equation}
	\nabla\mathcal{L}_{BM3D-MAP}(T) = \sum^{KM}_{i=1} 2 \, \mathbf{p}_i\, \mathbf{T}^V \left( 1 -\frac{y_i}{|\mathbf{p}_i\, \mathbf{T}^V|^2}\right)\mathbf{p}_i^{\dagger} + \mu_1 \left(\mathbf{T}^V - \mathbf{T}^V_{sparse} \right)
	\end{equation}
	Since \ref{equ:LMAP2} and \ref{equ:LMAP3} are non-convex functions, there is no guarantee that standard gradient descent converges to a global minimum. Recently, a non-convex algorithm for the generalized phase retrieval problem with Poisson noise was presented \cite{chen_solving_2015} that provably converges to a global minimum with suitable initialization. It introduces a iteration-dependent regularization on the gradients of the likelihood to remove terms which have a negative effect on the search direction. Namely, it introduces a truncation criterion 
	\begin{equation}
	\mathcal{E}^i(\mathbf{T}^V) = \left\lbrace \left|y_i - \left|\mathbf{p}_i \mathbf{T}^V\right|^2\right| \leq \frac{\alpha_h}{KM} 
	\norm{\mathbf{y} - \mathbf{I}}{1} \frac{\left|\mathbf{p}_i \mathbf{T}^V\right|}{\norm{\mathbf{T}^V}{2}} \right\rbrace
	\end{equation}
	that acts on the gradient of the likelihood and suppresses the influence of measurements that are too incompatible with the reconstruction. The regularized likelihood gradient is then 
	\begin{equation}
	\nabla\mathcal{L}_{\mathcal{E}^i}(\mathbf{T}^V) = \sum_{i \in \mathcal{E}^i(\mathbf{T}^V)}^{KM}\left[|\mathbf{p}_i\, \mathbf{T}^V|^2- y_i\, \mathrm{log}(|\mathbf{p}_i\, \mathbf{T}^V|^2)\right].
	\label{equ:LMAP4}
	\end{equation}
	We compute the next step using conjugate gradient descent \cite{shewchuk_introduction_1994,torch.optim}, since this lead to much faster convergence compared to the update procedure described in \cite{chen_solving_2015}.
	\subsubsection*{Initialization}
\begin{figure}[h!]
		\includegraphics[width=1\textwidth]{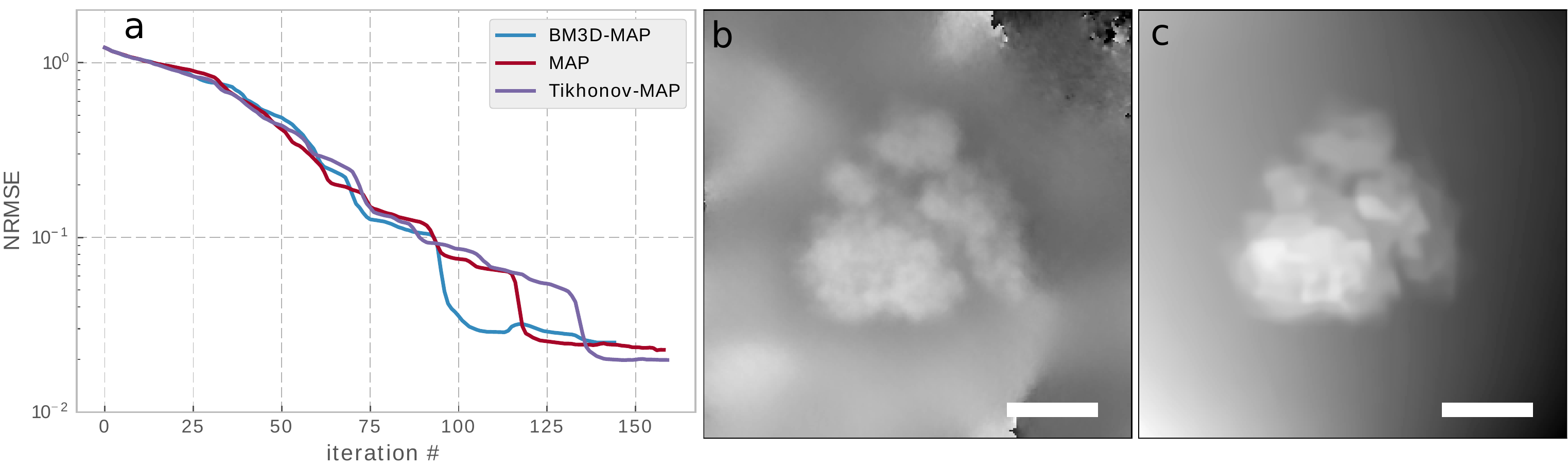}
		\caption{a) Convergence behavior of different gradient update rules b) Example for the transmission function initialization $T^0$ after 70 power iterations, for an electron dose of \SI{100}{\elementarycharge^-\per\angstrom^2}, intensities were truncated at the 80th percentile. c) $\mathbf{T}^V_{sparse}$ for human ribosome after 60 iterations of BM3D-MAP. Scale bar is \SI{10}{\nano\meter}.}
		\label{fig:algos}
	\end{figure}

	Truncated spectral initialization for ptychography was first proposed by Marchesini et al. \cite{Marchesin2014}, based on the notion that the highest intensities in the diffraction pattern carry the strongest phase information. They compute the phase of the largest eigenvector of the following hermitian operator:
	\begin{equation}
	\mathbf{1}_{|y_i|>\epsilon} \mathbf{F} \mathbf{Q}(\mathbf{Q}^{\dagger}\mathbf{Q})^{-1}\mathbf{Q}^{\dagger}\mathbf{F}^{\dagger}\mathbf{1}_{|y_i|>\epsilon}\, ,
	\end{equation}
	where $\mathbf{1}_{\mathbf{y}>\epsilon}$ is an indicator vector of the same dimension as $\mathbf{y}$ and $\epsilon$ is chosen so that the largest 20 percent of the intensities are allowed to contribute.
	The largest eigenvalue of a sparse hermitian matrix can be efficiently computed either with power iterations \cite{mises_praktische_1929}, or with the Arnoldi method \cite{lehoucq1998arpack}. In \cite{chen_solving_2015}, truncated spectral initialization with a truncation rule  with $\mathbf{1}_{|y_i|<\alpha_0^2\, \lambda_0^2}$ is used, with $\lambda_0=\sqrt{\sum_{i=1}^{KM} y_i}$. We  found the initialization of Marchesini et al. to produce better initializations for ptychographic datasets. This may be due to the nature of the ptychographic measurement vectors $\mathbf{p}_i$, which are different from the measurement vectors used in \cite{chen_solving_2015}. We leave further analysis of this matter for future research. It is also important to note that the truncated spectral initialization only produces visually correct initial phase to a dose of roughly \SI{100}{\elementarycharge^-\per\angstrom^2}. Fig. \ref{fig:algos} b) shows an example initialization for a dose of \SI{100}{\elementarycharge^-\per\angstrom^2}. For doses below this value, we initialized the transmission function with unity transmission and normal-distributed phase with mean \num{0.1}, \num{0.2}, and \num{0.3} for hemoglobin, 20S proteasome and ribosome respectively, and variance of \num{0.1}, which usually provides a lower starting error than the spectral initialization. With this sample-dependent random initialization we found no problem of convergence for all algorithms tested in this paper.
	\subsubsection*{Reconstruction parameters}
	All ptychography reconstructions were performed with a probe area overlap of \SI{75}{\percent} in real-space, where the probe area is defined by all pixels contributing more than \SI{1}{\percent} of the maximum intensity. This corresponds to a step size of roughly \SI{3}{\nano\meter}, depending on the probe used. For the reconstructions shown in Fig. \ref{fig:reconstructions} a total of 589 diffraction patterns were created using the random illumination. At a dose of \SI{20}{\elementarycharge^-\per\angstrom^2}, this corresponds to \num{5000} electrons per diffraction pattern.
	For the regularization parameters we performed a grid search evaluating the final NRMSE and found the values to be $\mathrm{\mu_1=4e-2}$, $\mathrm{\mu_2=1e-1}$. We choose the biorthogonal spline wavelet transform as the linear transform for BM3D as it achieves the best PSNR for high noise \cite{lebrun_analysis_2012}.
	\subsubsection*{Implementation Details}
	The algorithms presented in this paper were implemented with the torch scientific computing framework \cite{torch}. The gradient update routines were adapted from the optim package for torch \cite{torch.optim}. For efficient  computing on the graphics processing unit (GPU) with complex numbers, the zcutorch library for cuda was developed \cite{zcutorch}. hyperparameter optimization was done with the hypero \cite{hypero} package for torch. For BM3D denoising we use the C++ implementation \cite{lebrun_bm3d_????}.
	The code was run on an Intel i7-6700 processor with 32GB RAM and a NVidia Titan X GPU with 12GB RAM. The runtime for optimization with $\mathcal{L}_{MAP}$ was \SI{26}{\second}, and for optimization with $\mathcal{L}_{BM3D-MAP}$ \SI{35}{\second}, because the BM3D algorithm used here is not implemented on the GPU and the BM3D denoising is computationally more intensive.
	\section*{Conclusion}
	In this paper we have demonstrated via numerical experiments the possibility to retrieve high resolution electron transmission phase information of biological macromolecules using ptychography and Bayesian optimization. Using the methods presented in this paper, it should be possible to achieve resolutions below \SI{1}{\nano\meter} for true single particle imaging of large molecular complexes like human ribosome and a resolution around \SI{4}{\angstrom} with simple averaging of 15 datasets. We have given a detailed explanation of the optimization and initialization procedures used and have emphasized the importance of choosing an appropriate illumination function. We note that, while the high data redundancy in a ptychographic dataset empirically makes it experimentally very robust, there is much room for improvements in terms of measurement complexity. For the results presented here, the measurement dimension $\mathrm{KM}$ is larger than the problem dimension $\mathrm{N}$ by a factor of at least 30, while the theoretical limit for successful phase retrieval is $\mathrm{KM=4N}$ \cite{balan_signal_2006}. By reducing the number of measurements the variance of each individual measurement could be reduced, yielding an improved SNR in the reconstruction. Therefore the development of an optimized experimental scheme including design of the illumination function and scanning scheme is a promising direction of research and may enable significant improvement to the results presented here.\\
	We would like to point out two obstacles that one may have to overcome in the experimental realization of our results. Firstly the best results are to be expected when recoding zero-loss diffraction patterns with the use of an energy filter. The energy filter may introduce phase distortions into the diffraction patterns which may need to be accounted for in the reconstruction algorithm. One could achieve this by first reconstructing the incoming wavefront in characterization experiment without energy filter and then reconstructing the aberration induced by the energy filter with a fixed, known incoming wavefront. Secondly, although beam induced movements are expected to be reduced by a large amount due to spot-scanning, the remaining movement may cause problems in the reconstruction. Statistically stationary sample movements can be accounted for in the reconstruction algorithm \cite{pelz_--fly_2014,Clark2011}, but beam induced motions are likely to be non-stationary, and dedicated algorithms may need to be developed to account for it. Cryogenic ptychographic imaging of biological samples is also being developed in the X-ray sciences \cite{diaz_three-dimensional_2015}, and our results could equally be implemented there to improve the dose-effectiveness.
	Finally, the methods presented here may find application in electron phase imaging of radiation-sensitive samples under non-cryogenic conditions and the incorporation of Bayesian methods into in-focus ptychographic reconstruction procedures \cite{yang_enhanced_????,rodenburg_phase_1989} may provide similar gains in SNR as the ones discussed here while also keeping the analytical capabilities of traditional STEM imaging.
	\section*{Acknowledgements}
	This work was funded by the Max Planck Society through institutional support. P.M.P. acknowledges support from the International Max Planck Research School for Ultrafast Imaging \& Structural Dynamics.
	\section*{Author contributions statement}
	P.M.P., R.J.D.M. and R.B. conceived the idea. P.M.P and W.Q wrote the simulation code. P.M.P wrote the ptychographic reconstruction code. P.M.P wrote the paper with input and discussions from R.B, G.K, and R.J.D.M.
	\section*{Competing financial interests}
	The authors declare no competing financial interests.
	\begin {table}[h!]
	\begin{center}
		\begin{tabular}{|l|c|c|c|c|}
			\hline			
			Reference                                           & resolution  & \SI{}{\elementarycharge^-\per\angstrom^2} \\
			\hline
			D'Alfonso et al. \cite{dalfonso_dose-dependent_2016} & $\sim$\SI{1.5}{\angstrom}  & \num{3.98e4}\\ 
			Yang et al. \cite{yang_simultaneous_2016}           & atomic & \num{1.3e4}\\ 
			Putkunz et al.\cite{Putkunz2012}           &  $\sim$\SI{1}{\angstrom}   & \num{9.2e6}\\ 
			Humphry et al. \cite{Humphry2012} & $\sim$\SI{2.3}{\angstrom}    & \num{3.33e3} \\ 
			\hline  		
		\end{tabular}
		\caption{List of previously published electron ptychography experiments and used electron dose}\label{tab:title} 
	\end{center}
	\end {table}
	\bibliography{\jobname}

\end{document}